\documentclass[12pt]{article}
\usepackage{graphicx}
\usepackage{mathptmx,mathrsfs}
\usepackage{amsfonts,amsmath,amssymb,amsthm}
\usepackage{indentfirst}
\usepackage{cite}
\usepackage{hyperref}
\usepackage{enumitem}
\usepackage{bm,graphics,color}

\numberwithin{equation}{section}
\theoremstyle{definition}

\begin{document}

\title{BRST symmetry and fictitious parameters}

\author{A. A. Nogueira$^{1,2}$\thanks{andsogueira@hotmail.com}\; and B. M. Pimentel$^{3}$\thanks{pimentel@ift.unesp.br}\\
\textit{$^{1}${\small Instituto de Matem\'{a}tica e Computa\c{c}\~{a}o (IMC), Universidade Federal de Itajub\'a}}\\
\textit{\small Av. BPS 1303, Bairro Pinheirinho, CEP 37500-903, Itajub\'a, MG, Brazil}\\
\textit{$^{2}${\small Centro de Ci\^{e}ncias Naturais e
Humanas (CCNH), Universidade Federal do ABC}}\\
\textit{\small Av. dos Estados 5001, Bairro Santa Terezinha CEP 09210-580, Santo Andr\'{e}, SP, Brazil}\\
\textit{{$^{3}${\small Instituto de F\'{i}sica Te\'orica (IFT), Universidade Estadual Paulista}}}\\
\textit{\small Rua Dr. Bento Teobaldo Ferraz 271, Bloco II Barra Funda, CEP
01140-070 S\~ao Paulo, SP, Brazil}\\
}
\maketitle
\date{}

\begin{abstract}
Our goal in this work is to present the variational method of fictitious parameters and its connection with the BRST symmetry. Firstly we implement the method in QED at zero temperature and then we extend the analysis to GQED at finite temperature. As we will see the core of the study is the general statement in gauge theories at finite temperature, assigned by Tyutin work, that the physical degrees of freedom does not depend on the gauge choices, covariant or not, due to BRST symmetry.
\end{abstract}

\newpage

\section{Introduction}

One of the most important mathematical theorems already proven, among those which have guided the development of modern physics, are the theorems of Emmy Noether, where we realize the differential invariants by calculus of variations \cite{Noether}. In a physical language the implied symmetries in the classical equations of motion are synthesized in terms of an action, lagrangian and the consequent conserved quantities. Going forward the quantum equations of motion emerged from the quantization problem of a classical dynamics and it was first put in a formal language by Dirac \cite{WdW}, because he observed, with the correspondence principle, that the classical dynamics described in the phase space by the observable and Poisson brackets was associated to a quantum dynamics described in the Hilbert space by the operators and commutator{$\backslash$}anti-commutators. Later the existence of constraints in the hamiltonian dynamics led Dirac to extend its mechanical analysis of the phase space, defining the parenthesis of Dirac and the classification (first class{$\backslash$}second class), always being able to make connection to quantum dynamics with the correspondence principle\cite{Dirac1}. This is the first look at the connection between classical{$\backslash$}quantum dynamics. The second look begins in a Dirac study about the connection between a dynamics described in the configuration space and its resulting quantum dynamics. In this study we see the emergence of an object very important called transition amplitude \cite{Dirac2}. Feynman uses the idea of Dirac to formulate a way to describe the quantum lagrangian mechanics with the path integral formalism \cite{Feyn}. However we can see again the Noether's spirit in the elegant variational principle of quantum action by Schwinger functional formulation, utilizing as a guide the Heisenberg description \cite{Schwinger}. The breath of the Noether's theorems does not end, Matsubara and Fradkin with the matrix density of states and the principle of maximum entropy, include the thermodynamic equilibrium in the Schwinger formulation \cite{Mats}.

As we know the need to describe the interactions of nature along the lines of a relativistic dynamics led us to build a covariant language with a gauge symmetry \cite{Ut}. But this covariant language has more degrees of freedom then the physical ones and its necessary to impose constrains. The connections between classical and quantum physical systems with  constraints, in a functional formalism, was first formulated by Faddeev and later the ideas where extended by Senjanovic \cite{Fadd}. The quantization procedure of a gauge theory is only possible in the physical degrees of freedoms and thus we lost, in the process, the explicit covariance of the equations. In order to maintain the explicit covariance in quantum level Faddeev, Popov and DeWitt built a method where there is the appearance of the ghosts \cite{faddeev}. Despite in the Fadeev-Popov-DeWitt method we have not one action with gauge symmetry\footnote{The quantum gauge symmetry is maintained by the Ward-Takahashi identities \cite{WT}} there is a residual symmetry due to ghosts known as Bechi-Rouet-Stora-Tyutin (BRST) symmetry \cite{BRS,Tuytin}. The Tuytin work has a general statement that the physics does not depend on the gauge choices due to BRST symmetry, in other words, we see the variation of parameters and the use of Ward identities. Finally Batalin, Fradkin and Vilkovisky make an overview of the problem in view of covariant formalisms and its connection with BRST symmetry \cite{Fradk}, know as BRST quantization or BFV formalism. For more details about the covariant dynamics of systems with constraints and the quantization procedure \cite{Sunder}. The conection between the covariance and gauge symmetry can also be seen in the literature by the background field method which is a technique that allows us to fix the gauge and calculate the quantum effects without losing gauge invariance explicitly, in other words, the gauge symmetry is manifest throughout the calculation process in terms of the background field. The technique was started by DeWitte\cite{DWitt}. The main idea was to build a quantum theory of gravitation clearly covariant. The background field plays a key role, it is the field that keeps the covariance. But the technique only computed quantum contributions at one loop and the extension was necessary, attributed to t'Hooft \cite{Gthooft}. Weinberg \cite{Weinb} used the formalism in question to calculate the evolution of coupling constants by the effective action. In this method, although we construct effective action with explicit gauge invariance, the on-shell physics are independent of this fact\cite{LFA}. In non-Abelian gauge theories, there is a generalization of  Ward-Takahashi identities called the Slavnov-Taylor identities\cite{Slanov}. The geometrical interpretation of the Slavnov-Taylor identities is related to the BRST symmetry and the motions of the gauge fixing surface. Just to complete, in quantum gravity we see the gauge dependence of the one-loop effective action in DeWitt formalism together with the variation of the parameters and Ward-Slanov-Taylor identities\cite{Kallosh}. On the other hand, the gauge independence of the renormalized S-matrix has been proved in general gauge theories by \cite {Vonorov}. Notice that, nowadays the independence of on shell physics by a class of gauges choices can be seen by the variations of parameters and the Nielsen identities \cite{Nielsen}.

To be clear about the gauge fixing and its connection with BRST symmetry let us look at a particular case, the Maxwell electrodynamics (QED). As we know to quantize the theory using the Faddeev-Senjanovic method, the usual gauge chosen is the Coulomb gauge. Therefore we work with the true physical degrees of freedom but lost the explicit covariance. So in order to maintain the explicit covariance, we utilize the Faddeev-Popov-DeWitt trick and the gauge chosen is the Lorenz gauge. But we know that we can use other covariant gauge choices, such as the t'Hooft-Veltman covariant choice \cite{k2,Nash}, and there are many other choices of gauge in the literature, covariant or not. Naturally we think if these various choices of gauge do not contribute to the physical degrees of freedom there must be a symmetry behind this fact. In the present case, the different gauge choices are tied by the BRST symmetry, nicely demonstrated by Tuytin.

In the same form there is another electrodynamics in the literature, known as Podolsky generalized electrodynamics (GQED). As a small historical introduction let us say more about this kind of electrodynamics. When Ostrogradski built higher-order derivative Lagrangians in classical mechanics a new research field was opened \cite{Ostro}. Today we can put the set of higher-order theories as effective theories\footnote{The generalized electrodynamics, with  higher order derivatives, is unitary and stable \cite{Russos}.}. The main idea of this branch of higher-order derivatives is very simple; we construct additional higher-order terms in standard Lagrangians in such a way that it preserves the original symmetries of the problem\footnote{This can be seen in the generalization of Utijama's work \cite{Ut2}.}. The pioneers of higher-order theories in field theory were Bopp, Podolsky and Schwed \cite{Bop}. They proposed a generalized electrodynamics in an endeavor to get rid of the infinites on Maxwell electrodynamics, such as the electron self-energy and the vacuum polarization current. As a result, their expression for this new Lagrangian exhibits a free parameter or length that can be identified as the inverse of a mass, the Podolsky's mass $m_{p}$\footnote{The lower bound for the free parameter is $m_{p}\geq 350$ GeV \cite{Daniel}.}. The fact that the theory is finite in a classic point of view, will reflect in the same way in a quantum description. The theory has the quality of control the UV divergences in a sense closely related to the Pauli-Villars-Rayski regularization scheme.\cite{bufalo12, Rays}. So we describe the interaction in a covariant language with gauge symmetry and all the concepts discussed previously, about the covariant dynamics of systems with constraints and the quantization procedure with the inclusion of thermic effects, are applicable \cite{Boni1}. An important comment on how to fix the physical degrees of freedom in a covariant way should be done, although initially Podolsky had used the Lorenz condition

\begin{equation*}
\Omega \lbrack A]=\partial ^{\mu }A_{\mu }
\end{equation*}
to fix the physical degrees of freedom, after a rigorous study involving constraint analysis, we see from the classical point of view that this is not completely true, because we have a residual freedom when we utilize it. Explicitly, the Lorenz choice does not fix the gauge, it is not attainable (we do not recover the generalized Coulomb gauge from Lorenz gauge), and it is not preserved by the time evolution of the system \cite{galvao}. As a result, the natural way of fixing the degrees of freedom has become the generalized Lorenz condition

\begin{equation*}
\Omega \lbrack A]=(1+\frac{\square }{m_{p}^{2}})\partial ^{\mu }A_{\mu }
\end{equation*}
attainable, but this condition increases the order of the derivatives in the Lagrangian. On the other hand, there is other gauge condition known as no-mixing gauge \cite{pdif,AndGsdkp}

\begin{equation*}
\Omega \lbrack A]=\left( 1+\frac{\square }{m_{p}^{2}}\right) ^{\frac{1}{2}%
}\partial ^{\mu }A_{\mu }
\end{equation*}
that combines perfectly with the Podolsky theory, maintaining the order of the Lagrangian. But in the case of the no-mixing gauge condition it is necessary to contour a pseudo-differential structure. So the natural question is if the Lorenz condition, no-mixing and generalized Lorenz are tied by the BRST symmetry, despite of the peculiarities of each gauge choice \cite{AndTese}.

Therefore this work is devoted to the analysis of the BRST symmetry and its connection with gauge choices. In Sec.\ref{sec1} we established the variational method of fictitious parameters and the BRST symmetry in Maxwell electrodynamics, in which we used the metric signature $(+,-,-,-)$ for the Minkowski spacetime. In Sec.\ref{sec2} we extend the analysis to the Podolsky electrodynamics at thermodynamic equilibrium. In Sec.\ref{sec3} the authors present their final remarks and prospects.

\section{The BRST symmetry in QED and the fictitious parameters}

\label{sec1}

As it is known when we work with the covariant equations of motion in QED we utilize the Lorenz gauge to fix the physical degree of freedom. But there is other covariant gauge choice in the literature known as t'Hooft-Veltman choice \cite{k2,Nash}
\begin{equation}
\Omega=\partial_{\mu}A^{\mu}+gA_{\mu}A^{\mu},
\end{equation}
wherein its attainable is guaranteed by the Bell-Treiman transformations and $g$ is a free real parameter.

So let us analyse the BRST symmetry in the t'Hooft-Veltman gauge choice. Initially we start writing the transition amplitude with the Fadeev-Popov-DeWitt method

\begin{eqnarray}
\label{ampli2}
&&Z=\langle0||0\rangle=\int DA_{\mu}det[\Box+2gA^{\mu}\partial_{\mu}]exp[i\int d^{4}x {\cal L}]\cr\cr
&&{\cal L}=-\frac{1}{4}F^{\mu\nu}F_{\mu\nu}-\frac{[\partial_{\mu}A^{\mu}+gA_{\mu}A^{\mu}]^{2}}{2\xi},
\end{eqnarray}
where we can write $det[\Box+2g\partial_{\mu}A^{\mu}]$ in terms of ghosts fields

\begin{equation}
det[\Box+2g\partial_{\mu}A^{\mu}]=\int D\bar{c}Dc\exp\{-i\int d^{4}x\bar{c}[\Box+2gA^{\mu}\partial_{\mu}]c\}.
\end{equation}

Therefore,
\begin{eqnarray}
&&Z=\int DA_{\mu}D\bar{c}Dcexp[iS]\cr\cr
&&S=\int d^{4}x[-\frac{1}{4}F^{\mu\nu}F_{\mu\nu}-\frac{(\partial_{\mu}A^{\mu}+gA_{\mu}A^{\mu})^{2}}{2\xi}
-\bar{c}(\Box+2gA^{\mu}\partial_{\mu})c].
\end{eqnarray}
Note that in this gauge choice we have a fictitious interaction between the ghosts and the gauge field. Utilizing the Stueckelberg-Feynman phenomenology, the t'Hooft-Veltman gauge has as implications the vertex ghost-ghost-photon, triple and quartic interaction of photons in terms of the dummy parameter g. QED in this gauge is usually called pre-QCD due to phenomenological similarity of photon{$\rightarrow$}gluon vertices\footnote{QED in t'Hooft-Veltman gauge choice is a fantasmagoric QCD because we know that in QED we can take out the ghosts but in QCD not. There is, in some sense, a ficticious mimic between QED and QCD in this gauge.}.

Well, if the transition amplitude does not depend on g, we conclude that
\begin{eqnarray}
\label{BRST3}
\frac{\delta Z}{\delta g}&=&i\int DA_{\mu}D\bar{c}Dc\frac{\delta S}{\delta g}\exp[iS]\cr\cr
&=&i\int DA_{\mu}D\bar{c}Dc[\int d^{4}x(A_{\nu}[-\frac{1}{\xi}(\partial_{\mu}+gA_{\mu})A^{\mu}]A^{\nu}-2A_{\nu}\bar{c}\partial^{\nu}c)\exp[iS]=0,
\end{eqnarray}
and thus we have\footnote{From $Z$ with sources,
\begin{equation*}
Z[J_{\mu},\bar{\zeta},\zeta]=\int DA_{\mu}D\bar{c}Dc\exp[iS_{eff}];\quad S_{eff}=S+\int d^{4}x[J_{\mu}A^{\mu}+\bar{\zeta}c+\bar{c}\zeta]\,(\text{effective action}),
\end{equation*}
we can define the generator of connected Green's function $W$,
\begin{equation*}
iW=\ln Z;\quad\langle A^{\mu}\rangle=\frac{\delta W}{\delta J_{\mu}};\quad\langle c\rangle=\frac{\delta W}{\delta\bar{\zeta}};\quad\langle\bar{c}\rangle=-\frac{\delta W}{\delta\zeta},
\end{equation*}
and by a Legendre transformation on $W$ we define the generator of one particle irreducible amputed Green's function,
\begin{equation*}
\Gamma=W+\int d^{4}x[J_{\mu}\langle A^{\mu}\rangle+\bar{\zeta}\langle c\rangle+\langle\bar{c}\rangle\zeta].
\end{equation*}
}

\begin{eqnarray}
\label{BRST1}
&&\frac{\delta Z}{\delta g}=i\langle0|\int d^{4}x d^{4}y\delta^{4}(x-y)\{-\frac{1}{\xi}[\partial_{\mu}A^{\mu}(x)+gA_{\mu}(x)A^{\mu}(x)]A_{\nu}(y)A^{\nu}(y)+\cr\cr
&&+2\bar{c}(x)A_{\nu}(y)\partial^{\nu}c(y)\}|0\rangle=0,
\end{eqnarray}

\begin{equation}
\label{sameeq}
\langle0|-\frac{1}{\xi}[\partial_{\mu}A^{\mu}(x)+gA_{\mu}(x)A^{\mu}(x)]A_{\nu}(y)A^{\nu}(y)|0\rangle=\langle0|2\bar{c}(x)A_{\nu}(y)\partial^{\nu}c(y)|0\rangle.
\end{equation}
The previously equation associate the ghosts fields with photons fields, so it is relate to the no physical sector.

The BRST symmetry in the present case is given by the following fields transformations
\begin{eqnarray}
&&A_{\mu}\rightarrow A_{\mu}+\delta A_{\mu}\cr\cr
&&c\rightarrow c+\delta c\cr\cr
&&\bar{c}\rightarrow\bar{c}+\delta\bar{c},
\end{eqnarray}
and the variations are given by

\begin{eqnarray}
\label{tBRST3}
&&\delta A_{\mu}=\lambda\partial_{\mu}c \cr\cr
&&\delta c=0\cr\cr
&&\delta\bar{c}=-\frac{1}{\xi}\lambda(\partial_{\mu}+gA_{\mu})A^{\mu},
\end{eqnarray}
where $\lambda$ is a grassmannian parameter\footnote{The action $S$ with the gauge term and ghosts is invariant by the proposed BRST transformations.}.

So before the previously BRST fields transforms, the functional generator transforms as it follows

\begin{eqnarray}
&&Z'=Z+\delta_{BRST} Z\cr\cr
&&\delta_{BRST} Z=\int \delta_{BRST} (DA_{\mu}D\bar{c}Dc)\exp[iS_{eff}]+i\int DA_{\mu}D\bar{c}Dc \delta_{BRST} S_{eff}\exp[iS_{eff}],
\end{eqnarray}
on what

\begin{equation}
S_{eff}=S+\int d^{4}x[J_{\mu}A^{\mu}+\bar{\zeta}c+\bar{c}\zeta].
\end{equation}
Again, working with the measure of integration

\begin{equation}
DA_{\mu}D\bar{c}Dc\rightarrow J DA_{\mu}D\bar{c}Dc
\end{equation}

\begin{equation}
J=\left(\begin{array}{lll}
1& \lambda\partial_{\mu}\delta(x-y)& 0\\
-\frac{1}{\xi}\lambda(\partial_{\mu}+gA_{\mu})\delta(x-y)& 1& 0\\
0& 0& 1
\end{array}\right),
\end{equation}
we noticed that the Jacobian is equal to 1. On the other hand,

\begin{eqnarray}
\label{varZ}
\delta_{BRST} S_{eff}&=&\int d^{4}x[J_{\mu}\delta A^{\mu}+\bar{\zeta}\delta c+\delta\bar{c}\zeta]\cr\cr
&=&\lambda\int d^{4}x[J_{\mu}\partial^{\mu}c-\frac{1}{\xi}(\partial_{\mu}+gA_{\mu})A^{\mu}\zeta].
\end{eqnarray}
In this case, imposing the BRST symmetry on the functional $Z$ we found the equation\footnote{Eq. (\ref{varZ}), derived from the BRST symmetry $\delta_{BRST}Z=0$, could be written as follows \cite{Shapiro}
\begin{equation*}
\int d^{4}z\{J_{\mu}(z)\partial^{\mu}c(z)-\frac{1}{\xi}[\partial_{\mu}A^{\mu}(z)+gA_{\mu}(z)A^{\mu}(z)]\zeta(z)\}\,Z[J_{\mu},\bar{\zeta},\zeta]=0,
\end{equation*}
where the fields and sources are seen as operators acting on $Z=\exp iW$,
\begin{equation*}
A_{\mu}(z)=\frac{\delta}{i\delta J^{\mu}(z)};\quad c(z)=\frac{\delta}{i\delta\bar{\zeta}(z)};\quad\bar{c}(z)=-\frac{\delta}{i\delta \zeta(z)}.
\end{equation*}}

\begin{equation}
\delta_{BRST} Z=\int DA_{\mu}D\bar{c}Dc\delta_{BRST}S_{eff}\exp[iS_{eff}]=0.
\end{equation}
Replacing $\delta_{BRST}S_{eff}$ in the previous equation and dividing it by Z we have the functional average \cite{Kallosh, Vonorov}

\begin{equation}
\langle\int
d^{4}z\{J_{\mu}(z)\partial^{\mu}c(z)-\frac{1}{\xi}[\partial_{\mu}A^{\mu}(z)+gA_{\mu}(z)A^{\mu}(z)]\zeta(z)\}\rangle=0,
\end{equation}
in which we see a Slanov-Taylor identity\cite{Slanov}.\footnote{We could interpret the Slanov-Taylor average equation as follows,
\begin{equation*}
\delta_{BRST}W=-i\delta_{BRST}\ln Z=\frac{\int DA_{\mu}D\bar{c}Dc\delta_{BRST}S_{eff}\exp[iS_{eff}]}{\int DA_{\mu}D\bar{c}Dc\exp[iS_{eff}]}=\langle\delta_{BRST}S_{eff}\rangle.
\end{equation*}
}

Multiplying the Slanov-Taylor identity by Z and applying the operator $\frac{\delta^{3} }{\delta J_{\nu}(y)\delta J^{\nu}(y)\delta\zeta(x)}$ we have, taking the sources equal to zero, the relation between the ghosts and photons fields\footnote{For reasons of completeness, due to certain divergence in the community, we present the steps to obtain eq. (\ref{BRST4}) to criticism of the full audience. When we multiplying the Slanov-Taylor average identity by Z we have the equation,

\begin{equation*}
\int DA_{\mu}D\bar{c}Dc\int d^{4}z\{J_{\mu}(z)\partial^{\mu}c(z)-\frac{1}{\xi}[\partial_{\mu}A^{\mu}(z)
+gA_{\mu}(z)A^{\mu}(z)]\zeta(z)\}\exp[iS_{eff}]=0.
\end{equation*}
Differentiating the previous equation with respect to $\zeta(x)$, taking into account its anticommutativity:
\begin{eqnarray*}
&&\int DA_{\mu}D\bar{c}Dc\big\{-i\bar{c}(x)\int d^{4}z\big(J_{\mu}(z)\partial^{\mu}c(z)-\frac{1}{\xi}[\partial_{\mu}A^{\mu}(z)
+gA_{\mu}(z)A^{\mu}(z)]\zeta(z)\big)+\cr\cr
&&-\frac{1}{\xi}[\partial_{\mu}A^{\mu}(x)
+gA_{\mu}(x)A^{\mu}(x)]\big\}\exp[iS_{eff}]=0.
\end{eqnarray*}
Then differentiate again with respect to $J^{\nu}(y)$:
\begin{eqnarray*}
&&\int DA_{\mu}D\bar{c}Dc\big\{\bar{c}(x)A^{\nu}(y)\int d^{4}z\big(J_{\mu}(z)\partial^{\mu}c(z)-\frac{1}{\xi}[\partial_{\mu}A^{\mu}(z)
+gA_{\mu}(z)A^{\mu}(z)]\zeta(z)\big)+\cr\cr
&&-i\bar{c}(x)\partial^{\mu}c(y)-\frac{1}{\xi}[\partial_{\mu}A^{\mu}(x)
+gA_{\mu}(x)A^{\mu}(x)]A^{\nu}(y)\big\}\exp[iS_{eff}]=0.
\end{eqnarray*}
Finally, differentiate the above equation with respect to $J_{\nu}(y)$ and, set the sources equal to zero, we have eq. (\ref{BRST4}) in the explicitly form:
\begin{equation*}
\int DA_{\mu}D\bar{c}Dc\{2\bar{c}(x)A^{\nu}(y)\partial_{\nu}c(y)+\frac{1}{\xi} [\partial_{\mu}A^{\mu}(x)+gA_{\mu}(x)A^{\mu}(x)]A^{\nu}(y)A_{\nu}(y)\}\exp[iS]=0.
\end{equation*}
}

\begin{equation}
\label{BRST4}
\langle0|-\frac{1}{\xi} [\partial_{\mu}A^{\mu}(x)+gA_{\mu}(x)A^{\mu}(x)]A^{\nu}(y)A_{\nu}(y)|0\rangle=\langle0| 2\bar{c}(x)A^{\nu}(y)\partial_{\nu}c(y)|0\rangle.
\end{equation}
Therefore the variation of $Z$ in eq. (\ref{BRST1}) has as consequence eq .(\ref{sameeq}), that is equal to eq. (\ref{BRST4}). So the condition that $Z$ does not depend of the fictitious parameter $g$ in eq. (\ref{BRST3}), in the sense that the variation under this parameter is equal to zero, lead us to think that the independence of transition amplitude due to the dummy parameter $g$ is associated with the BRST symmetry.

We conclude that the different covariat choices of gauge
\begin{equation}
\label{c1}
\Omega=\partial_{\mu}A^{\mu}\;(\text{Lorenz});\quad\Omega=\partial_{\mu}A^{\mu}+gA_{\mu}A^{\mu}\;(\text{t'Hooft-Veltman})
\end{equation}
are possible due the BRST symmetry because the parameter involved, $g$, does not contribute to the transition amplitude. Now the independence of the physics by the t'Hooft parameter $g$ was see, for example, in the cancelation when we compute the first correction to photon propagator in the momentum representation by Feynman rules\cite{Nash}.

\section{The BRST symmetry in GQED at thermodynamics equilibrium}

\label{sec2}

Previously we have discussed the connection between the covariant gauge choices and BRST symmetry in Maxwell quantum electrodynamics.
At this moment we are able to apply the same set of ideas in a wider context, studying the connection between the covariant gauge choices and BRST symmetry in Podolsky quantum electrodynamics at thermodynamics equilibrium \cite{Boni1}.

\subsection{General covariant gauge choice}
We started the method with the following Lagrangian density in thermodynamic equilibrium

\begin{eqnarray}
\label{Naka}
&&{\cal\hat{L}}_{N}=\frac{1}{4}\hat{F}_{\mu\nu}\hat{F}_{\mu\nu}+\frac{1}{2m^{2}_{p}}\partial_{\mu}\hat{F}_{\mu\lambda}\partial_{\theta}\hat{F}_{\theta\lambda}+
\frac{1}{2}\{\hat{B},G[\hat{A}]\}+\frac{\xi}{2}\hat{B}^{2}\cr\cr
&&\hat{F}_{\mu\nu}=\partial_{\mu}\hat{A}_{\nu}-\partial_{\nu}\hat{A}_{\mu}
\end{eqnarray}
where $\hat{B}$ is the auxiliary field of Nakanishi and $G[\hat{A}]$ is an operator of gauge choice \cite{Nakani}. We will make the following general covariant choice\footnote{The d'Alembert in the Euclidean is of the form $\triangle=-\partial_{\mu}\partial_{\mu}$.}
\begin{equation}
\label{c2}
G[\hat{A}]=(\frac{\triangle}{m^{2}_{p}}+1)^{\epsilon}\partial_{\mu}\hat{A}_{\mu}.
\end{equation}
wherein

\begin{equation}
\epsilon=0\;(\text{Lorenz});\quad\epsilon=\frac{1}{2}\;(\text{No-mixing});\quad\epsilon=1\;(\text{Generalized Lorenz}).
\end{equation}
To find the quantum equations of motion in thermodynamic equilibrium we use the variational principle of Schwinger\footnote{The integral in the action is define as $\int_{\beta} d^{4}x= \int^{\beta}_{0} d\tau \int_{V} d^{3}x$, where $\tau$ is the inverse of the thermal energy and $V$ is the physical volume.}

\begin{eqnarray}
&&\hat{S}=\int_{\beta} d^{4}x{\cal\hat{L}}_{N},\cr\cr
&&\delta\hat{S}=\hat{0}.
\end{eqnarray}
Therefore before applying the changes of the fields in the previous action

\begin{eqnarray}
&&\hat{A}_{\mu}\rightarrow\hat{A}_{\mu}+\delta\hat{A}_{\mu}\cr\cr
&&\hat{B}\rightarrow\hat{B}+\delta\hat{B}
\end{eqnarray}
with the relations,

\begin{eqnarray}
&&\frac{1}{4}\hat{F}_{\mu\nu}\hat{F}_{\mu\nu}\rightarrow-\frac{1}{2}\hat{A}_{\mu}(\delta_{\mu\nu}\triangle+\partial_{\mu}\partial_{\nu})\hat{A}_{\nu}\cr\cr
&&\frac{1}{2m^{2}_{p}}\partial_{\mu}\hat{F}_{\mu\lambda}\partial_{\theta}\hat{F}_{\theta\lambda}\rightarrow
-\frac{1}{2}\hat{A}_{\mu}(\delta_{\mu\nu}\triangle+\partial_{\mu}\partial_{\nu})\frac{\triangle}{{m^{2}_{p}}}\hat{A}_{\nu},
\end{eqnarray}
remembering  that the surface terms do not contribute to the action, the equations of motion are given by

\begin{eqnarray}
&&(\frac{\triangle}{m^{2}_{p}}+1)(\delta_{\mu\nu}\triangle+\partial_{\mu}\partial_{\nu})\hat{A}_{\nu}
-(\frac{\triangle}{m^{2}_{p}}+1)^{\epsilon}\partial_{\mu}\hat{B}=\hat{0}\cr\cr
&&\hat{B}=-\frac{1}{\xi}(\frac{\triangle}{m^{2}_{p}}+1)^{\epsilon}\partial_{\nu}\hat{A}_{\nu}.
\end{eqnarray}
So we have,

\begin{equation}
-(\frac{\triangle}{m^{2}_{p}}+1)\{\delta_{\mu\nu}\triangle+
[1-\frac{1}{\xi}(\frac{\triangle}{m^{2}_{p}}+1)^{2\epsilon-1}]\partial_{\mu}\partial_{\nu}\}\hat{A}_{\nu}=\hat{0}
\end{equation}
and then we define que differential Podolsky operator
\begin{equation}
P^{(m^{2}_{p},\xi,\epsilon)}_{\mu\nu}=-(\frac{\triangle}{m^{2}_{p}}+1)\{\delta_{\mu\nu}\triangle+
[1-\frac{1}{\xi}(\frac{\triangle}{m^{2}_{p}}+1)^{2\epsilon-1}]\partial_{\mu}\partial_{\nu}\}.
\end{equation}

Although we fix the gauge, there is a residual symmetry in the theory. This can be observed by making a gauge transformation into eq. (\ref{Naka})

\begin{eqnarray}
&&\hat{A}_{\mu}\rightarrow\hat{A}_{\mu}+\partial_{\mu}\hat{\alpha}\cr\cr
&&\hat{B}\rightarrow\hat{B}\cr\cr
&&\delta{\cal\hat{L}}_{N}=\frac{1}{2}\{\hat{B},(\frac{\triangle}{m^{2}_{p}}+1)^{\epsilon}\triangle\hat{\alpha}\}.
\end{eqnarray}
In this case in order to have symmetry

\begin{equation}
\delta{\cal\hat{L}}_{N}=\hat{0}\Rightarrow(\frac{\triangle}{m^{2}_{p}}+1)^{\epsilon}\triangle\hat{\alpha}=\hat{0}.
\end{equation}
Implementing the previous condition by a quantum Lagrange multiplier $\hat{\lambda}$ in thermodynamic equilibrium

\begin{equation}
{\cal\hat{L}}={\cal\hat{L}}_{N}+\hat{\lambda}(\frac{\triangle}{m^{2}_{p}}+1)^{\epsilon}\triangle\hat{\alpha}.
\end{equation}
we can write a general form of the Lagrangian density

\begin{equation}
{\cal\hat{L}}={\cal\hat{L}}_{N}+i\hat{\bar{c}}(\frac{\triangle}{m^{2}_{p}}+1)^{\epsilon}\triangle\hat{c}.
\end{equation}
where the Lagrange multiplier $\hat{\lambda}$ and the gauge parameter $\hat{\alpha}$ are naturally identifies as the grassmannians ghosts fields.

\subsection{The partition function}
The above analysis leads us to define directly the partition function \cite{Fra}

\begin{eqnarray}
&&Z=Z_{0}\int DA_{\mu} D\bar{c}Dcexp[-S],\cr\cr
&&S=\int_{\beta} d^{4}x[\frac{1}{2}A_{\mu}P^{(m^{2}_{p},\xi,\epsilon)}_{\mu\nu}A_{\nu}-i\bar{c}(\frac{\triangle}{m^{2}_{p}}+1)^{\epsilon}\triangle c].
\end{eqnarray}
In fact, remembering the functional representation of the determinants, we conclude that

\begin{equation}
Z=Z_{0}\det[P^{(m^{2}_{p},\xi,\epsilon)}]^{-\frac{1}{2}}\det[(\frac{\triangle}{m^{2}_{p}}+1)^{\epsilon}\triangle]
\end{equation}
Now given a differential operator M we know that

\begin{eqnarray}
&&M=A\delta_{\mu\nu}+B\partial_{\mu}\partial_{\nu},\cr\cr
&&\det[M]=A^{4}-A^{3}B\triangle.
\end{eqnarray}
In the present case $M=P^{(m^{2}_{p},\xi,\epsilon)}$

\begin{eqnarray}
&&A=-(\frac{\triangle}{m^{2}_{p}}+1)\triangle,\cr\cr
&&B=[1-\frac{1}{\xi}(\frac{\triangle}{m^{2}_{p}}+1)^{2\epsilon-1}].
\end{eqnarray}
Therefore,

\begin{equation}
\det[P^{(m^{2}_{p},\xi,\epsilon)}]=\det[(\frac{\triangle}{m^{2}_{p}}+1)\triangle]^{4}\det[\frac{1}{\xi}(\frac{\triangle}{m^{2}_{p}}+1)^{2\epsilon-1}].
\end{equation}
So we have

\begin{eqnarray}
\label{simetria}
Z&=&Z_{0}\det[(\frac{\triangle}{m^{2}_{p}}+1)\triangle]^{-2}\det[\frac{1}{\xi}(\frac{\triangle}{m^{2}_{p}}+1)^{2\epsilon-1}]^{-\frac{1}{2}}
\det[(\frac{\triangle}{m^{2}_{p}}+1)^{\epsilon}\triangle],\cr\cr
&=&Z_{0}\det[m^{2}_{p}]^{-\frac{3}{2}}\det[\frac{1}{\xi}]^{-\frac{1}{2}}\det[1+\frac{\triangle}{m^{2}_{p}}]^{-\frac{3}{2}}\det[\triangle]^{-1}.
\end{eqnarray}

We conclude then, by the  last equation, that the partition function does not depend on the dummy parameter $\epsilon$ and thus it does not depend on the class of choice defined in eq. (\ref{c2}). In this way as the object partition function is intrinsically related to the physical degree of freedom the gauge choices does not affect them. As we know the physical degrees of freedom (Maxwell+Proca, d=2+3) of Podolsky theory are seen in this way

\begin{eqnarray}
&&\det[\triangle]^{-1}\quad(Maxwell)\;d=2,\cr\cr
&&\det[1+\frac{\triangle}{m^{2}_{p}}]^{-\frac{3}{2}}\quad(Proca)\;d=3.
\end{eqnarray}

\subsection{The BRST symmetry}
At this moment we will apply all the study of BRST symmetry in a particular gauge condition, know as no-mixing gauge. As we saw the partition function in this gauge is given by

\begin{equation}
Z=Z_{0}\det[P^{(m^{2}_{p},\xi,\epsilon)}]^{-\frac{1}{2}}\det[(\frac{\triangle}{m^{2}_{p}}+1)^{\epsilon}\triangle],\quad(\epsilon=\frac{1}{2}).
\end{equation}
Note that the ghost sector in the no-mixing gauge condition is a pseudo-differential structure. To avoid this problem

\begin{equation}
\det[(\frac{\triangle}{m^{2}_{p}}+1)^{\epsilon}\triangle]=\frac{\det[(\frac{\triangle}{m^{2}_{p}}+1)^{\epsilon+\frac{1}{2}}\triangle]}
{\det[(\frac{\triangle}{m^{2}_{p}}+1)]^{\frac{1}{2}}}.
\end{equation}
Therefore the partition function is written as

\begin{eqnarray}
\label{ampli3}
&&\tilde{Z}=Z_{0}\int DA_{\mu} D\bar{c}Dc D\phi exp[-\tilde{S}],\cr\cr
&&\tilde{S}=\int_{\beta} d^{4}x[\frac{1}{2}A_{\mu}P^{(m^{2}_{p},\xi,\epsilon)}_{\mu\nu}A_{\nu}-i\bar{c}(\frac{\triangle}{m^{2}_{p}}+1)^{\epsilon+\frac{1}{2}}\triangle c+\frac{1}{2}\phi(\frac{\triangle}{m^{2}_{p}}+1)\phi].
\end{eqnarray}
where $\bar{c}$, $c$ are grassmann fields and $\phi$ is a real scalar field. Immediately we can see that

\begin{eqnarray}
\label{fantasma}
\tilde{Z}&=&Z_{0}\det[P^{(m^{2}_{p},\xi,\epsilon)}]^{-\frac{1}{2}}\det[(\frac{\triangle}{m^{2}_{p}}+1)^{\epsilon+\frac{1}{2}}\triangle]
\det[(\frac{\triangle}{m^{2}_{p}}+1)]^{-\frac{1}{2}}\cr\cr
&=&Z_{0}\det[P^{(m^{2}_{p},\xi,\epsilon)}]^{-\frac{1}{2}}\det[(\frac{\triangle}{m^{2}_{p}}+1)^{\epsilon}\triangle]\cr\cr
&=&Z
\end{eqnarray}
and the interpretation is given saying that the scalar field eats the degrees of freedom of the grassmann field maintaining the physical degrees of freedom of Podosky theory. Note that the ghost sector has now two sectors, fermionic and bosonic, and when we take the limit $\epsilon\rightarrow\frac{1}{2}$ there is no pseudo-differential structure.

Inserting the sources in view of to build an effective Schwinger's action
\begin{eqnarray}
\label{fungen}
&&\tilde{Z}=Z_{0}\int DA_{\mu} D\bar{c}Dc D\phi exp[-\tilde{S}_{eff}]\cr\cr
&&\tilde{S}_{eff}=\tilde{S}+\int d^{4}x[J_{\mu}A_{\mu}+\bar{\zeta}c+\bar{c}\zeta+J\phi],
\end{eqnarray}
and varying $\tilde{Z}$ with respect to $\epsilon$

\begin{equation}
\delta\tilde{Z}=\frac{\delta\tilde{Z}}{\delta\epsilon}\delta\epsilon.
\end{equation}
We have then

\begin{eqnarray}
\frac{\delta\tilde{Z}}{\delta\epsilon}&=&-Z_{0}\int DA_{\mu} D\bar{c}Dc D\phi \frac{\delta\tilde{S}_{eff}}{\delta\epsilon}exp[-\tilde{S}_{eff}]\cr\cr
&=&-Z_{0}\int DA_{\mu} D\bar{c}Dc D\phi\{\int_{\beta}
d^{4}x[\frac{1}{2}A_{\mu}\frac{\delta P^{(m^{2}_{p},\xi,\epsilon)}_{\mu\nu}}{\delta\epsilon}A_{\nu}-i\bar{c}
\frac{\delta(\frac{\triangle}{m^{2}_{p}}+1)^{\epsilon+\frac{1}{2}}}{\delta\epsilon}\triangle c]\}\times\cr\cr
&\times&exp[-\tilde{S}_{eff}],
\end{eqnarray}
and so
\begin{eqnarray}
&&\frac{\delta\tilde{Z}}{\delta\epsilon}=-Z_{0}\int DA_{\mu} D\bar{c}Dc D\phi\{\int_{\beta}
d^{4}x d^{4}y\delta^{4}(x-y)[\frac{\delta P^{(m^{2}_{p},\xi,\epsilon)}_{\mu\nu}}{\delta\epsilon}\frac{1}{2}A_{\mu}(x)A_{\nu}(y)+\cr\cr
&&-i\frac{\delta(\frac{\triangle}{m^{2}_{p}}+1)^{\epsilon+\frac{1}{2}}}{\delta\epsilon}\triangle\bar{c}(x)c(y)]\}
\exp[-\tilde{S}_{eff}].
\end{eqnarray}
Imposing that the partition function doesn't depend of $\epsilon$

\begin{equation}
\frac{\delta\tilde{Z}}{\delta\epsilon}=0\Rightarrow\frac{\delta P^{(m^{2}_{p},\xi,\epsilon)}_{\mu\nu}}{\delta\epsilon}\frac{1}{2}\langle A_{\mu}(x)A_{\nu}(y)\rangle=i\frac{\delta(\frac{\triangle}{m^{2}_{p}}+1)^{\epsilon+\frac{1}{2}}}{\delta\epsilon}\triangle\langle\bar{c}(x)c(y)\rangle.
\end{equation}
With the identities

\begin{eqnarray}
&&\frac{\delta P^{(m^{2}_{p},\xi,\epsilon)}_{\mu\nu}}{\delta\epsilon}=\frac{2}{\xi}(\frac{\triangle}{m^{2}_{p}}+1)^{2\epsilon-1}\ln(\frac{\triangle}{m^{2}_{p}}+1)\partial_{\mu}\partial_{\nu}\}\cr\cr
&&\frac{\delta(\frac{\triangle}{m^{2}_{p}}+1)^{\epsilon+\frac{1}{2}}}{\delta\epsilon}=
(\frac{\triangle}{m^{2}_{p}}+1)^{\epsilon+\frac{1}{2}}\ln(\frac{\triangle}{m^{2}_{p}}+1),
\end{eqnarray}
the last equation is written as follows

\begin{equation}
\frac{1}{\xi}(\frac{\triangle}{m^{2}_{p}}+1)^{\epsilon-\frac{1}{2}}\partial_{\mu}\partial_{\nu}\langle A_{\mu}(x)A_{\nu}(y)\rangle=i\triangle\langle\bar{c}(x)c(y)\rangle.
\end{equation}
So we have the thermal average equation
\begin{equation}
\label{BRST5}
\frac{1}{\xi}\partial_{\mu}\langle A_{\mu}(x)A_{\nu}(y)\rangle=i\partial_{\nu}\langle\bar{c}(x)c(y)\rangle\quad(\epsilon=\frac{1}{2}).
\end{equation}
As we can see, the last equation is consequence of the non-mixing gauge choice and the imposition that the partition function doesn't depend of $\epsilon$ in the variational method, other gauge choice could modify it. Now it remains to show that the last equation is the same as that one from BRST symmetry.

Firstly note that the BRST symmetry in the present case is given by the following fields transformations
\begin{eqnarray}
&&A_{\mu}\rightarrow A_{\mu}+\delta A_{\mu}\cr\cr
&&c\rightarrow c+\delta c\cr\cr
&&\bar{c}\rightarrow\bar{c}+\delta\bar{c}\cr\cr
&&\phi\rightarrow\phi+\delta\phi,
\end{eqnarray}
where the variations are given by

\begin{eqnarray}
\label{BRST6}
&&\delta A_{\mu}=i\lambda\partial_{\mu}c \cr\cr
&&\delta c=0\cr\cr
&&\delta\bar{c}=-\frac{1}{\xi}\lambda\partial_{\mu}A^{\mu}\cr\cr
&&\delta\phi=0,
\end{eqnarray}
with $\lambda$ a grassmannian parameter, change the lagrangian density as follows

\begin{eqnarray}
{\cal \tilde{L}}&=&\frac{1}{\xi}A_{\mu}[(1+\frac{\triangle}{m^{2}_{p}})\partial_{\mu}\partial_{\nu}] \delta A_{\nu}-i\delta\bar{c}(1+\frac{\triangle}{m^{2}_{p}})\triangle c\cr\cr
&=&\frac{i}{\xi}\partial_{\mu}[i\lambda A_{\mu}(1+\frac{\triangle}{m^{2}_{p}})\triangle c]
\end{eqnarray}
and then the invariance of the action $\tilde{S}$ is guaranteed because total derivative does not contribute to action. Finally imposing the BRST symmetry defined in eq. (\ref{BRST6}) on the thermodynamic functional generator in eq. (\ref{fungen}), we found that
\begin{equation}
\int d^{4}z[iJ_{\mu}(z)\partial_{\mu}c(z)-\frac{1}{\xi}\partial_{\mu}A_{\mu}(z)\zeta(z) ]\,Z[J_{\mu},\bar{\zeta},\zeta,J]=0.
\end{equation}
Applying the operator $\frac{1}{Z}\frac{\delta^{2} }{\delta J_{\nu}(y)\delta\zeta(x)}$ in the above equation, taking the sources equal to zero, we have eq. (\ref{BRST5}), an equation that relates the longitudinal sector of
the photon to the ghost sector. The physical sector is the transverse.

As we can see by eq. (\ref{simetria}), the partition function $Z$ does not depend on the dummy
parameter $\epsilon$. So the variation of the partition function with respect to $\epsilon$ will be zero and as a result of this operation we have eq. (\ref{BRST5}), an equation from BRST symmetry. This fact lead us to think that the independence of the transition amplitude due to the dummy parameter $\epsilon$ is associated with the BRST symmetry.

\section{Conclusion and final remarks}

\label{sec3}

In this paper we have analyzed the link between the covariant gauge choices, fictitious parameters and BRST symmetry. Implicitly, we again experienced the beautiful, elegant and simple variational ideas of Emmy Noether in a covariant quantum problem with the inclusion of the thermodynamic equilibrium.

Firstly in Maxwell electrodynamics we see by the variational method of fictitious parameters that the transition amplitude does not depend on the fictitious parameter. So the Lorenz and t'Hooft-Veltman gauges describes the same physical degrees of freedom.

In the same way in Podolsky electrodynamics at finite temperature the Lorenz, non-mixing and generalized Lorenz choice describes the same physical degrees of freedom because the partition function does not depend on these covariant choices, seen in eq. (\ref{simetria}). By analyzing the BRST symmetry we chose the non-mixing gauge, due to the problem involving a pseudo-differential structure. This problem is solved in a peculiar way, synthesized in the eq. (\ref{fantasma}), on which we have an interesting interpretation. Note that now the ghost has two sectors, one grassmanian and other scalar, and the scalar field eats the degrees of freedom of the grassmann field maintaining the physical degrees of freedom of Podosky theory.

As final remarks let us talk about some aspects. Since the beginning, the BRST symmetry guarantee the renormalization program of a gauge theory, relating certain ultraviolet divergences (UV) that appear in radiative corrections in the regularization process \cite {brs}. Following the same line of reasoning, nowadays there are consequences when we choose the gauge, associated with the UV divergences in generalized electrodynamics \cite{bufalo2}. For example, the Lorenz condition generates certain UV divergences associated with radiative corrections of the fermion propagator and vertex in GQED that does not appear in generalized Lorenz or no-mixing gauges. We say then that the Lorenz gauge generates UV divergences. But we saw earlier that the physics does not depend on the gauge choices, so we apply them to calculate physical results. As we can see, the Lorenz gauge is applied in the works \cite{FGA} where we see stationary physical interactions between charges, dipoles, conductors and fine solenoids. On the other hand the no-mixing gauge is applied to calculate the Casimir effect in Podolsky electrodynamics and we see that this gauge choice is easier to handle than Lorenz or generalized Lorenz, because of the shape of the generalized photon propagator \cite{AndRussia}. We believe that for better understanding of the connection between a quantum dynamics, described by the physical degree of freedoms in the radiation gauge (Coulomb), and its correspondent covariant dynamic (Lorenz, no-mixing, generalized Lorenz), the ghosts and the BRST symmetry should be include, studying the theory in a general context of BFV formalism \cite{Bufalo3}. In BFV formalism the method of Faddeev-Popov-DeWitte is elucidated and questions about attainability of gauges are discussed clearly. This work complements previous studies of the authors on covariant quantum dynamics \cite{AndGsdkp, Proceedings} and opens the door to a more complete approach, involving the Matsubara-Fradkin formalism of quantum covariant dynamics in equilibrium \cite{Contin}.

\section{Acknowledgement}

A. A. Nogueira thanks Capes and (PNPD/Capes-UFABC) for partial support and B. M. Pimentel thanks CNPq for partial support.

\end{document}